\documentclass[oneside, 11pt, reqno ]{amsart}
\usepackage[dvips, bottom=1.3in, top=1.2in ]{geometry}

\usepackage{amsmath, amssymb}
\usepackage{enumerate,color,comment}
\usepackage{pstricks, pst-node, pst-tree}
\usepackage{stmaryrd}

\usepackage[sort&compress, square, numbers]{natbib}
\numberwithin{equation}{section}
\theoremstyle{plain}                
\newtheorem{theorem}{Theorem}[section]

\newtheorem{proposition}[theorem]{Proposition}

\theoremstyle{definition}           
\newtheorem{definition}[theorem]{Definition}
\newtheorem{example}[theorem]{Example}

\theoremstyle{remark}
\newtheorem{remark}[theorem]{Remark}

\renewcommand{\cite}[1]{\citep{#1}}
\newcommand{\citeand}[2]{\citetext{\citealp{#1} and \citealp{#2}}}

\DeclareMathOperator*\esssup{esssup}

\newcommand{\tot}{\tfrac{1}{2}} 
\newcommand{\abs}[1]{\left| #1 \right|} 
\newcommand{\set}[1]{\left\{#1\right\}} 
\newcommand{\sets}[2]{\set{#1\,:\,#2}} 

\newcommand{\ft}[2]{#1\dots#2} 
\renewcommand{\ft}[2]{#1,\dots,#2}

\newcommand{\prf}[1]{ ( #1 )_{t\in [0,T]}} 
\newcommand{\prfi}[1]{ ( #1 )_{t\in [0,\infty)}} 

\newcommand{\RN}[2]{\frac{d#1}{d#2}}

\providecommand{\R}{} \renewcommand{\R}{{\mathbb R}}

\newcommand{\N}{{\mathbb N}}
\newcommand{\PP}{{\mathbb P}}
\newcommand{\QQ}{{\mathbb Q}}

\newcommand{\EE}{{\mathbb E}}
\newcommand{\FF}{{\mathcal F}}

\newcommand{\BB}{{\mathcal B}}

\newcommand{\MM}{{\mathcal M}}
\newcommand{\NN}{{\mathcal N}}

\newcommand{\FFF}{{\mathbb F}}

\newcommand{\EN}{{\mathcal E}}

\renewcommand{\AA}{{\mathcal A}}

\newcommand{\eps}{\varepsilon}
\newcommand{\ld}{\lambda}

\newcommand{\el}{{\mathbb L}} 

\newcommand{\linf}{\el^{\infty}}

\newcommand{\tpds}[1]{\tfrac{\partial}{\partial #1}}

\newcommand{\define}[1]{{\em #1}}

\newcommand{\mnotesign}{$\bigstar$}
\newcommand{\mnote}[1]{\textcolor{red}{\mnotesign}\marginpar[]{\footnotesize
    \textcolor{red}{\mnotesign} #1}}

\newcommand\tB{{\tilde{B}}}

\newcommand\sL{{\mathcal L}}

\newcommand{\Xxp}{X^{x,\pi}}

\newcommand{\PPt}{\PP|_{\FF_t}}
\newcommand{\PPT}{\PP|_{\FF_T}}

\newcommand{\QQt}{\QQ^{(t)}}

\newcommand{\MMt}{\MM_t}
\newcommand{\MMs}{\MM_s}
\newcommand{\MMi}{\MM_{\infty}}

\newcommand{\AAb}{\AA_{\mathrm{bd}}}
\newcommand{\ii}{{\textstyle \int_0^{\infty} \pi_s\, dS_s}}
\renewcommand{\it}{{\textstyle \int_0^{t} \pi_s\, dS_s}}
\newcommand{\iT}{{\textstyle \int_0^{T} \pi_s\, dS_s}}

\newcommand{\LL}{{\mathbb L}}
\renewcommand{\define}[1]{{\bf #1}}

\newcommand{\hrho}{\hat{\rho}}

\newcommand{\Cs}{C^{(s)}}
\newcommand{\Csr}{C^{(r)}}

\newcommand{\sqt}[1]{\{#1\}_{t\in \N}}
\newcommand{\sqtz}[1]{\{#1\}_{t\in \N_0}}

\reversemarginpar
\renewcommand{\mnotesign}{$\bigstar$}
\renewcommand{\mnote}[1]{\textcolor{red}
  {\mnotesign}\marginpar[{\footnotesize
    \textcolor{red}{\mnotesign} #1}]{}}
\renewcommand{\mnote}[1]{ {\bf Note:} {\em #1} }
\renewcommand{\mnotesign}{{\color{red} $\bigstar$}}
\renewcommand{\mnote}[1]{\mnotesign \marginpar{ \mnotesign #1}}
\renewcommand{\mnote}[1]{[ \mnotesign: {\em #1 }]}

\title{Maturity-independent risk measures}

\begin{document}

\author{Thaleia Zariphopoulou}

\address{Thaleia Zariphopoulou, Departments of Mathematics and
  Information, Risk and Operations Management,
  The University of Texas at Austin, 1 University Station, C1200,
  Austin, TX, USA} \email{ zariphop@math.utexas.edu}

\author{ Gordan \v{Z}itkovi\'{c} }

\address{ Gordan \v Zitkovi\' c, Department of Mathematics, The
  University of Texas at Austin, 1 University Station, C1200, Austin,
  TX, USA} \email{gordanz@math.utexas.edu}

\thanks{ We thank the participants of the ``Workshop on Financial
  Engineering and Actuarial Mathematics'', University of Michigan, May
  2007, the meeting on ``Stochastic Analysis in Finance and Insurance'',
  Oberwolfach, January 
  2008 and ``AMS Sectional Meeting'', Baton Rouge, March  2008 for
  engaging discussions and insightful comments.  Both authors
  acknowledge partial support from the National Science Foundation
  (NSF Grants: DMS-0091946, NSF-RTG-0636586 and DMS-FRG-0456118, and
  NSF Grant: DMS-0706947, respectively).  }

\date{\today}%

\begin{abstract}
  The new notion of maturity-independent risk measures is introduced
  and contrasted with the existing risk measurement concepts. It is
  shown, by means of two examples, one set on a finite probability
  space and the other in a diffusion framework, that, surprisingly,
  some of the widely utilized risk measures cannot be used to build
  maturity-independent counterparts. We construct a large class of
  maturity-independent risk measures and give representative examples
  in both continuous- and discrete-time financial models.
\end{abstract}


 \maketitle

\section{Introduction}

The abstract notion of a risk measure appeared first in
\cite{ArtDelEbeHea97} and \cite{ArtDelEbeHea99}.  The simple axioms
set forth in \cite{ArtDelEbeHea99} opened a venue for a rich field of
research that shows no signs of fatigue.  The main reason for such
success is the fundamental need for quantification and measurement of
risk. While the initial impetus came from the requirements of the
financial and insurance industries, applications in a wide range of
situations, together with a mathematical tractability and elegance of
this theory, have promoted risk measurement to an independent field of
interest and research. The early cornerstones include (but are not
limited to) 
\citeand{FolSch02, FolSch02b}{FriRos02}; see,
also, \cite{FolSch04} for more information.

The first notions of risk measures were all static, meaning that the
time of measurement, as well as the time of resolution (maturity,
expiry) of the risk were fixed. Soon afterwards, however, dynamic and
conditional risk measures started to appear (see
\citeand{ArtDelEbeHeaKu03, Rie04, CheDelKup04,  DetSca05,
CheDelKup05, CheDelKup06}{RooSchEng05}, as well the book \cite{FolSch04}).

Despite all the recent work in this wide area, there is still a number
of theoretical, as well as practical, questions left unanswered. The
one we focus on in the present paper deals with the problem one faces
when the maturity (horizon, expiration date, etc.) associated with a
particular risky position is not fixed.  We take the view that the
mechanism used to measure the risk content of a certain random
variable should not depend on any a priory choice of the measurement
horizon.  This is, for example, the case in complete financial
markets. Indeed, consider for simplicity the Samuelson (Black-Scholes)
market model with zero interest rate and the procedure one would
follow to price a contingent claim therein. The fundamental theorem of
asset pricing tells us to simply compute the expectation of the
discounted claim under the unique martingale measure. There is no
explicit mention of the maturity date of the contingent claim in this
algorithm, or, for that matter, any other prespecified horizon.
Letting the claim's payoff stay unexercised for any amount of time
after its expiry would not change its arbitrage-free price in any way.

It is exactly this property that, in our opinion, has not received
sufficient attention in the literature. As one of the fundamental
properties clearly exhibited under market completeness, it should be
shared by any workable risk measurement and pricing procedure in
arbitrary incomplete markets.

The incorporation of the maturity-independence property described
above into the existing framework of risk measurement has been guided
by the principle of minimal impact: we strove to keep new axioms as
similar as possible to the existing ones for convex risk measures, and to
implement only minimally needed changes. This led us to the
realization that it is the domain of the risk measure that
inadvertently dictates the use of a specific time horizon, and if we
replace it by a more general domain, the maturity-independence would
follow. Thus, our axioms are identical to the axioms of a
replication-invariant convex risk measure, except for the choice of
the domain which is {\em not} a subspace of a function space on
$\FF_T$, for some fixed time horizon $T$.

In addition to the novel axiom pertinent to maturity independence, a
link to the notion of forward performance processes, recently proposed
by M.~Musiela and the first author (see \cite{MusZar03, MusZar05,
  MusZar06a, MusZar06, MusZar07}) is established.  Indeed, focusing on
the exponential case, it is shown that every forward performance
process can be used to create an example of a maturity-independent
risk measure. On one hand, this connection provides a useful and
simple tool for (a non-trivial task of) constructing
maturity-independent risk measures. On the other hand, we hope that it
would give a firm decision-theoretic foundation to the theory of
forward performances.

We start off by introducing the financial model, trading and
no-arbitrage conditions, and recalling some well-known facts about
risk measures. In section 3, we introduce the notion of a
maturity-independent risk measure, argue for its feasibility and
relevance, and give first examples. We also show, via two simple
examples, that a na\"ive approach to the construction of
maturity-independent risk measures can fail. Section 4 opens with the
notion of a performance random field and goes on to describe the
important class of forward performance processes.  These objects are,
in turn, used to produce a class of maturity-independent risk measures
which we call {\em forward entropic risk measures}.  Finally, several
special cases of these measures are mentioned and interpreted in
section 5, and an independent example, set in a binomial-type
incomplete financial model, is presented.

\section{Generalities on the financial market model and risk measures}
\label{sec:general}
\subsection{Market Set-up, No-Arbitrage Conditions  and Admissible Portfolios}
\subsubsection{The  Model.}
Let $(\Omega, \FF,\FFF,\PP)$ be a complete probability space, with the
filtration $\FFF=\prfi{\FF_t}$ generated by a $d$-dimensional Brownian
motion $\prfi{W_t}=\prfi{W^1_t,\dots, W^d_t}$ (and augmented by the
$\PP$-null sets).  The evolution of the prices of risky assets is
modeled by an It\^ o-process $\prfi{S_t}=\prfi{S^1_t,\dots, S^k_t}$ of
the form
\begin{equation}%
  \label{equ:stock-price}
    \begin{split}
dS^i_t=S^i_t 
\Big( 
\mu^i_t\, dt+ \sum_{j=1}^d \sigma^{ji}_t\, dW^j_t
\Big),
    \end{split}
\end{equation}
for $t\geq 0$, $i=\ft{1}{k}$ and $j=\ft{1}{d}$, where the processes
$\prfi{\mu^i_t}$ and $\prfi{\sigma^{ji}_t}$, are $\FFF$-progressively
measurable and uniformly bounded by a deterministic constant. The
requirement of uniform boundedness can be replaced by a much less
stringent one, but we choose not to pursue such a generalization for
the sake of transparency.

We postulate the existence of a $d$-dimensional
progressively-measurable process $\prfi{\lambda_t}$ such that
\begin{equation}
    \label{equ:lambda}
    \begin{split}
      \sum_{j=1}^d \sigma^{ji}_t \lambda^j_t= \mu^i_t,\ i=\ft{1}{k},\
      t\geq 0, \text{ a.s.}
    \end{split}
\end{equation}
While such $\ld$ does not need to be unique, we will always work with
the
representative chosen in a minimal way, namely, 
\begin{equation}%
\label{equ:for-theta}
    \begin{split}
\lambda^j_t = \sum_{i=1}^k (\sigma^+)^{ji}_t \mu^i_t,\ t\geq
0,\text{ a.s.},
    \end{split}
\end{equation}
where $\sigma^+$ denotes the Moore-Penrose pseudo-inverse of the matrix
$\sigma$. Effectively, our choice of the market price of risk (vector)
process
$\lambda$ amounts to the solution of the matrix equation $\sigma
\ld=\mu$ with
the minimal Euclidean norm. We assume, in addition, that 
each component 
of the process $\ld^j$, just
like $\sigma^{ji}$ and $\mu^i$, is uniformly bounded by a
(deterministic) constant.  

The existence of a liquid risk-free asset $S^0$ is also postulated.
As usual, we quote all asset-prices in the units of
$S^0$. Operationally, this amounts to the simplifying assumption
$S^0_t= 1$, $t\geq 0$, which will hold throughout.

\subsubsection{Portfolio processes}
\label{sss:portfolio-proc} 
A $k$-dimensional $\FFF$-progressive process
$\pi=\prfi{\ft{\pi^1_t}{\pi^k_t}}$ is called a \define{portfolio
  (process)} if $\sum_{i=1}^k \int_0^{t} (\pi^i_s)^2\, ds<\infty$,
a.s., for all $t\geq 0$.  A portfolio $\pi$ is called \define{admissible} if there exists
a constant $a>0$ (possibly depending on $\pi$, but not on the state of
the world) such that the \define{gains process}
$X^{\pi}=\prfi{X^{\pi}_t}$, defined as
\[X^{\pi}_t=\it =\sum_{i=1}^k \int_0^t \pi^i_s\, dS^i_s, \ t\geq 0,\]
is bounded from below by $-a$, i.e., $ X^{\pi}_t\geq -a,\text{ for all
  $t\geq 0$, a.s. }$ The set of all portfolio processes $\pi$ whose
gains processes $X^{\pi}$ are admissible is denoted by $\AA$.  For
technical reasons, which will be clear shortly, we introduce the set
$\AAb$ of all portfolio processes $\pi$ whose gains process $X^{\pi}$
is uniformly bounded from above, as well as from below, i.e.,
$\AAb=\AA\cap (-\AA)=\sets{\pi\in \AA}{-\pi\in \AA}$.

\subsubsection{No Free Lunch with Vanishing Risk} 
The natural assumption of no arbitrage is routinely replaced in the
literature by the slightly stronger, but still economically feasible,
assumption of {\em no free lunch with vanishing risk} (NFLVR).  It was
shown in the seminal paper \cite{DelSch94} that, when postulated on
finite time-intervals $[0,t]$, $t\in (0,\infty)$, NFLVR is equivalent
to the following statement: for each $t\geq 0$, there exists a
probability measure $\QQt$, defined on $\FF_t$, with the following
properties:
\begin{enumerate}
\item $\QQt\sim \PPt$, where $\PPt$ is the restriction of the
  probability measure $\PP$ to $\FF_t$, and
\item the stock-price process $S$ is a $\QQt$-local martingale, when
  restricted to the interval $[0,t]$.
\end{enumerate}
It is well-known that, under the assumptions we imposed on the
coefficient processes $\mu$ and $\sigma$, the condition of NFLVR, and,
thus, the equivalent statement above, are automatically satisfied on
finite  intervals $[0,t]$, $t\in (0,\infty)$.  Therefore, for
$t\geq 0$, the set of all measures $\QQt$ with the above properties is
non-empty.  We will denote this set by $\MMt^e$.
\subsubsection{Closed market models}
\label{sss:closed}
It is immediate that, for $0\leq s<t$, we have the following relation
\begin{equation}
    \nonumber 
    \begin{split}
      \MMs^e=\sets{\QQt|_{\FF_s}}{\QQt\in\MMt^e}.
    \end{split}
\end{equation}
The restriction map turns the family $\prfi{\MMt^e}$
into an inversely directed system. In general, such a system will {\em
  not} have an inverse limit, i.e., there will exist {\em no} set
$\MM^e_{\infty}$ with the property that
$\MMt^e=\sets{\QQ|_{\FF_t}}{\QQ\in\MMi^e}$, for all $t$. In other
words, even though the market may admit no arbitrage (NFLVR) on any
finite interval $[0,t]$, arbitrage opportunities might arise if we
allow the trading horizon to be arbitrarily long.  In order to
differentiate those cases, we introduce the notion of a closed market
model:
\begin{definition}
\label{def:closed}
A market model $\prfi{S_t}$ is said to be \define{closed} if there
exists a set $\MMi^e$ of probability measures $\QQ\sim\PP$ such that,
for every $t\geq 0$, $\QQt\in\MMt^e$ if and only if
$\QQt=\QQ|_{\FF_t}$ for some $\QQ\in\MMi^e$.
\end{definition}
\begin{remark}
\label{rem:not-closed}
Most market models used in practice are not closed. The simplest
example is Samuelson's model, where the filtration is generated by a
single Brownian motion $\prfi{W_t}$, and the price of the risky asset
satisfies $dS_t=S_t(\mu\, dt+\sigma\, dW_t)$, for some constants
$\mu\in\R$, $\sigma>0$. For $t\geq 0$, the only element in $\MMt^e$
corresponds to a Girsanov transformation.  However, as $t\to\infty$,
this transformation becomes ``more and more singular'' with respect
to $\PPt$, and no $\QQ$ as in Definition \ref{def:closed} can be found
(see \cite{KarShr91}, Remark on p.~193).
\end{remark}
\subsection{Convex risk measures}
\subsubsection{Axioms of convex risk measures}
One of the main reasons for the wide use and general acceptance of the
theory of risk measures lies in its axiomatic nature. Only the most
fundamental  traits of an economic agent, such as risk aversion, are
encoded parsimoniously into the axioms of risk measures. The resulting
theory is nevertheless rich and relevant to the financial
practice. The pioneering notion of a coherent risk measure (see
\cite{ArtDelEbeHea99}) has, soon after its conception, been replaced by
a very similar, but more flexible, notion of a convex risk measure
(introduced in \citeand{Hea00, FolSch02, FriRos02}{HeaKu04}):
\begin{definition}  
\label{def:risk-meas}
A functional $\rho$ mapping $\linf(\Omega,\FF,\PP)$ into $\R$ is
called a \define{convex risk measure} if, for all $f,g\in \linf$, we
have
\begin{align*}
  (1) \qquad & \rho(f)\leq 0 \text{ if } f\geq 0,\text{ a.s.};
  & \text{\sf (anti-positivity)}\\
  (2) \qquad & \rho(f-m)=\rho(f)+m,\ m\in\R;
  & \text{\sf (cash-translativity)}\\
  (3) \qquad & \rho(\ld f + (1-\ld) g)\leq \ld
  \rho(f)+(1-\ld)\rho(g),\ \ld\in [0,1].  & \text{\sf (convexity)}
\end{align*}
\end{definition}

\subsubsection{Replication invariance}
The idea that two risky positions which differ only by a quantity
replicable in the market at no cost, should have the same risk content
has appeared very soon after the notion of a risk measure has been
applied to the study of financial markets.  In order to expand on this
tenet, let us, temporarily, pick an arbitrary time $T>0$, and suppose
that we are dealing with a finite-horizon financial market
$\prf{S_t}$, where all finite-horizon analogues of the assumptions and
definitions above hold. In such a situation, the investors will trade
in the market in order to reduce the overall risk of the terminal
position, as measured by the risk measure $\rho$ defined on
$\linf(\FF_T)$.  In other words, the combination of the financial
market and the risk measure $\rho$ will give rise to a new risk
measure, denoted herein by $\rho(\cdot; T)$, given by
\[ \rho(f;T)=\inf_{\pi\in\AAb} \rho\left(f+\iT\right).\] We will use the
$T$-notation to stress the dependence of this risk measure on the
specific maturity date.  In addition to the Axioms (1)-(3) from
Definition \ref{def:risk-meas}, the functional $\rho(\cdot;T)$
satisfies the following property:
\begin{align*}
  (4)\qquad & \rho(f;T)=\rho(f+\iT;T)\text{ for all }
  f\in\linf(\FF_T), \pi\in\AAb.
& \text{\sf (replication invariance)}
\end{align*}
\begin{definition} 
\label{def:m-c-risk-meas}
A mapping $\rho(\cdot; T):\linf(\Omega,\FF_T,\PP|_{\FF_T})\to\R$ is
called a \define{re\-pli\-ca\-ti\-on-invariant convex risk measure} if
it satisfies axioms (1)-(3) of Definition \ref{def:risk-meas} and
(4) above.
\end{definition}
The notion of replication invariance was introduced in
\cite{FolSch02}, and further developed and generalized in
\cite{FriSca06}. An accessible discussion of coherent and convex risk
measures, as well as the notion of  replication invariance, can be
found in chapter 4 of \cite{FolSch04}. 
\begin{remark}\ 
  \begin{enumerate}
  \item
  When the market model is complete, the restrictions imposed by 
  adding  the replication invariance axiom will necessarily force
  any replication-invariant risk measure to coincide with the
  replication price functional (the ``Black-Scholes price''). It is
  only in the setting of incomplete markets that the interplay between
  risk measurement and trading in the market produces a non-trivial
  theory.
\item It may seem somewhat counterintuitive at the first glance that a
  replication-invariant risk measure should assign the same risk
  content to the constant $S_0$ as to the random variable $S_T$ (where
  $\prf{S_t}$ is a price process of a traded risky asset). The
  resolution can be found in the fact that the risk contained in $S_T$
  is virtual since it can be hedged away completely in the financial
  market. Replication-invariant measures are, however,  typically not
  law-invariant, i.e., there are random variables with the same
  $\PP$-distribution as $S_T$ whose risk content is possibly much larger.
\end{enumerate}

\end{remark}
The following examples of (maturity-specific) replication-invariant
convex risk measures are very well known (see \cite{FolSch04}). 
We use them as test cases
for the notion of maturity in\-de\-pen\-den\-ce introduced in
Definition \ref{def:horiz-indep}, below. One can easily show that all
of them satisfy axioms (1)-(4).

\medskip
\begin{example}\ 
\label{exa:replication-invariant}
\begin{enumerate}
\item {\bf Super-hedging.}\ \ For $f\in\linf(\FF_T)$, let $\hrho(f;T)$
  be the super-hedging price of $f$, i.e.,
  \[ \hrho(f;T)=\inf\sets{m\in\R}{\exists\, \pi\in\AAb,\ \iT\geq
    m+f,\text{ a.s.}}.\] The risk measure $\hrho(\cdot;T)$ is extremal
  in the sense that, for each replication-invariant convex risk measure
  $\rho(\cdot;T)$, we have $\hrho(f;T)\geq \rho(f;T)$, for all
  $f\in\linf(\FF_T)$.
\item {\bf Entropic risk measures.}\ For $f\in\linf(\FF_T)$, the
  entropic risk measure $\rho(f;T)$, with risk aversion coefficient
  $\gamma>0$, is defined as the unique solution $\rho\in\R$ to the
  indifference-pricing equation
\begin{equation}
\label{equ:def-entr}
   \begin{split}
&\sup_{\pi\in\AAb} \EE\left[
  -\exp\left(-\gamma\left(x+\rho+f+\iT\right)\right) \right]=
\\
&\sup_{\pi\in\AAb} \EE\left[-\exp\left(-\gamma(x+\iT)\right)\right],\ x\in\R.
   \end{split}
\end{equation}
The value $\rho(-f;T)$ at the negative $-f$ of $f$ is also known as the
\define{exponential indifference price} $\nu(f;T)$ of $f$.
The measure $\rho(\cdot;T)$ admits a simple dual representation
\begin{equation}%
  \label{equ:El-Karoui}
  \begin{split}
    \rho(f;T)=\sup_{\QQ\in\MM^e_T} \Big(
    \EE^{\QQ}[-f]-\tfrac{1}{\gamma}H(\QQ|\PP; T) \Big),
    \end{split}
  \end{equation}
  where the \define{ relative entropy $H(\QQ|\PP;T)$ of
    $\QQ\in\MM^e_T$ with respect to $\PP$} is given by
  \[ H(\QQ|\PP;T)=\EE^{\QQ}\left[
    \ln\left(\RN{\QQ}{(\PPT)}\right)\right]\in [0,\infty].\]
\item {\bf General replication-in\-va\-ri\-ant risk measures.} 
Under appropriate topological regularity
  conditions  replication-in\-va\-ri\-ant convex risk measure
  $\rho(\cdot;T):\linf(\FF_T)\to\R$ admits the following dual
  representation
  \begin{equation}%
\label{equ:dual-char}
    \begin{split}
      \rho(f;T)=\sup_{\QQ\in\MM_T^e}\Big( \EE^{\QQ}[-f]-\alpha(\QQ)
      \Big),
    \end{split}
\end{equation}
for some convex penalty function $\alpha:\MM_T^e\to [0,\infty]$. See
Theorem 17, p.~445 in \cite{FolSch02} for the proof in the 
discrete-time case. The
proof in our setting is similar.  
\end{enumerate}
\end{example}

\section{Maturity-independent risk measures}

\subsection{The need for maturity independence}
\label{sse:hi}
The classical notion of a convex risk measure, as well as its
replication-invariant specialization, is inextricably linked to a
specific maturity date with respect to which risk measurement is
taking place while ignoring all other time instances. On the other
hand, a fundamental property of financial markets is that they
facilitate transfers of wealth among different time points as well as
between different states of the world. The notion of replication
invariance, introduced above, abstracts the latter property and ties
it to the decision-theoretic notion of a convex risk measure. The
former property, however, has not yet been incorporated into the risk
measurement framework in the same manner in the existing
literature. One of the goals herein is to do exactly this. We, then, 
pose and
address the following question: \begin{quotation}
``{\em 
Is there a class of risk measures that are not constructed in
reference to a specific time instance  and can be, thus,
used to measure the risk content of claims of all (arbitrary)
maturities?}''
\end{quotation}
Equivalently, we wish to avoid the case when two versions of the same
risk measure (differing only on  the choice of the maturity date)
give different risk values to the same contingent claim\footnote{
\label{rem:time-impatience}
One could object to the above  reasoning by pointing out
that different maturities {\em should} give rise to different
risk assessments due to the effect of 
time impatience. In response, we take a view that the market is
efficient in the sense that all time impatience is already incorporated
in the investment possibilities present in it. 
 More specifically, we remind the reader that the
assumption that $S^0\equiv 1$ effectively means that all contingent
claims are quoted in terms of time-$0$ currency. One can easily extend
the theory presented here to the more general case where the 
time-value of money is modeled explicitly. We feel, however, 
 that such a
generalization would only obscure the central issue herein  and render the
present paper less accessible. 
}.

\smallskip

Before we proceed with formal definitions, let us recall  some of the
fundamental properties of the arbitrage-free 
pricing (``Black-Scholes'') functional,
$\rho_{BS}$, in the context of a complete financial market. For a
``regular-enough'' 
contingent claim $f$, the value $\rho_{BS}(f)$ is defined as the 
 capital needed at inscription to replicate it perfectly. 
The
functional $\rho_{BS}$  satisfies the axioms of convex risk measures 
and  is replication-invariant. Moreover, it is per se unaffected by
 the expiration date of the generic claim $f$. 

When markets are incomplete, a much more interesting set of phenomena
occurs, as there is no canonical (``Black-Scholes'') pricing mechanism.  
We shall see  that, interestingly,  some traditional and
widely used risk measures 
are \emph{not} maturity-independent. 
In other words, under these measures, 
indifference prices of the 
same contingent claim, but calculated in terms of two distinct
 maturities will, in general, {\em differ}. 

\subsection{Definition of maturity independence}
Let  $\LL$ denote 
the set of all bounded random variables
with finite maturities, i.e.,  
\[ \LL=\cup_{t\geq 0} \linf(\FF_t).\] The set $\LL$ will serve as a
natural domain for the class of risk measures we propose in the
sequel. Note that $\LL$ contains all $\FF_t$-measurable bounded
contingent claims, for all times $t\geq 0$, but it avoids the (potentially
pathological) cases of random variables in $\linf(\FF_{\tau})$, where
$\tau$ is a finite, but possibly unbounded,  stopping time.  

We are now ready to define the class of
maturity-independent risk measures.  With a slight
abuse of notation, we still use the symbol $\rho$. 
 In contrast to
their maturity-dependent counterparts $\rho(\cdot;
T)$, however, all maturity-specific notation has vanished.

\begin{definition}
\label{def:horiz-indep}
  A functional $\rho:\LL\to\R$ is called a \define{maturity-independent
    convex risk measure} if it has the following properties for all
  $f,g\in\LL$, and $\ld\in [0,1]$:
\begin{equation}
    \nonumber 
    \begin{aligned}
      (1)\quad & \rho(f)\leq 0,\ \forall\, f\geq 0, 
     &\text{\sf (anti-positivity)}\\
      (2)\quad & \rho(\ld f+(1-\ld) g)\leq \ld\rho(f)+(1-\ld)\rho(g),
     &\text{\sf (convexity)}\\
      (3)\quad & \rho(f-m)=\rho(f)+m,\ \forall\,  m\in\R,\text{ and } 
     &\text{\sf (cash-translativity)}\\
      (4)\quad & \text{for \emph{all} }t\geq 0,\text{ and }\pi\in\AAb,\
     \rho(f+\it)=\rho(f).
     \hspace{-8ex}& \\ & 
     &\text{\sf (replication and maturity independence)}
    \end{aligned}
\end{equation}
\end{definition}
We note that the properties which  differentiate the maturity-independent risk
measures from the existing notions are the {\em choice of the domain}
$\LL$ on the one hand, 
and the validity of axiom (4) for {\em all} maturities $t\geq 0$ on the other.
\subsection{Motivational  examples}
We start off our investigation of maturity-independent risk measures 
by giving three examples - one of an extremal such
risk measure, 
one of a class of maturity-independent risk measures for closed markets,
and one in which the maturity independence property fails.

\subsubsection{Super-hedging prices}
\label{sss:super-hedging} 
The simplest example of a maturity-independent risk measure is the
{super-hedging price} function $\hat{\rho}:\LL\to\R$ given by
\[ \hat{\rho}(f)=\inf\sets{m\in\R}{\exists\,\pi\in \AAb, \
  m+\ii \geq f, \text{ a.s.}}.
\]
It is easy to see that it satisfies  all axioms in
Definition \ref{def:horiz-indep}. As in the
maturity-dependent case, $\hat{\rho}$ has the  extremal
 property
$\hat{\rho}(f)\geq \rho(f)$,  for any $f\in\LL$ and any maturity-independent
risk measure $\rho$.
\subsubsection{The case of closed markets}
\label{sss:closed-example}
The dual characterization (\ref{equ:dual-char}) of
replication-invariant risk measures for finite maturities can be used
to construct maturity-independent risk measures when the market model
is closed (see Definition \ref{def:closed} and paragraph
\ref{sss:closed} for notation and terminology). Indeed, let
$\alpha:\MMi^e\to [0,\infty]$ be a proper function (i.e., satisfying
$\alpha(\QQ)<\infty$, for at least one $\QQ\in\MMi^e$.) It is not
difficult to check that the functional $\rho:\LL\to\R$, defined by
\begin{equation}
    \nonumber 
    \begin{split}
\rho(f)=\sup_{\QQ\in\MMi^e} \Big( \EE^{\QQ}[-f]-\alpha(\QQ)\Big),
    \end{split}
\end{equation}
is a maturity-independent risk measure. We have already seen that many
market models used in practice are not closed. The natural
construction used above will clearly not be applicable in those cases
and, thus,  an entirely different approach will be needed.

\subsubsection{Risk measures lacking maturity independence}
\label{sss:no-hor-ind}

It is tempting to assume that a 
ma\-tu\-ri\-ty-independent risk measure
$\rho$ can always be constructed by identifying a 
maturity date $t$
associated with a contingent claim $f$, and setting
$\rho(f)=\rho(f;t)$, for some replication-invariant risk measure
$\rho(\cdot;t)$. As shown in the following two examples, this construction
will not always be possible even if we restrict our attention 
to the well-explored class of
entropic risk measures.
Both examples are based on the entropic risk measure (see Example
\ref{exa:replication-invariant} (2)).  Note that the first example is not
entirely set in the framework described in  section 2, but the
reader will easily make all required (formal) modifications. \\

\noindent\begin{minipage}{0.66\textwidth}
\paragraph{\bf a) A non-compliance example on a finite probability space.}
We present a simple two-period example in which 
entropic risk measurement 
gives
different results for the same, time-$1$-measurable contingent claim
$f$, when  considered at time $1$ and time $2$. The market
structure is described by the simple tree in Figure 1, where the
(physical) probability of each of the branches leaving the initial
node is $\tfrac{1}{3}$, and the conditional probabilities of the two
contingencies (leading to $S_4$ and $S_5$) 
after the node $S_3$ are equal to $\tfrac{1}{3}$ and
$\tfrac{2}{3}$, respectively. One can implement the described
situation on a $4$-element probability space $\Omega=\set{\omega_1,
  \omega_2, \omega_3, \omega_4}$, as in Figure 1, with
$\PP[\omega_1]=\PP[\omega_2]=1/3$, $\PP[\omega_3]=1/9$ and
$\PP[\omega_4]=2/9$.
\end{minipage}
\hspace{0.01\textwidth}
\begin{minipage}{0.3\textwidth}
\begin{center}
\footnotesize
\psset{labelsep=2pt, tnpos=a, radius=2pt, treefit=loose}
\pstree[treemode=R,levelsep=10ex]{\TC~{$S_0$}}{
  \pstree{\TC~{$S_1$}}{\TC~{$S_1$}~[tnpos=r]{$\omega_1$}}
  \pstree{\TC~{$S_2$}}{\TC~{$S_2$}~[tnpos=r]{$\omega_2$}}
  \pstree{\TC~{$S_3$}}{\TC~{$S_4$}~[tnpos=r]{$\omega_3$} 
                             \TC~{$S_5$}~[tnpos=r]{$\omega_4$} }
}

\ \\[1ex]

{\bf Figure 1.}\ The market tree \\[0.1ex]
\end{center}
\end{minipage}

\smallskip

\noindent 

  There are two financial
instruments: a riskless bond $S^0\equiv 1$, and a stock $S=S^1$ whose price is
denoted by $S_0,\dots, S_5$ for various nodes of the information tree,
such that the following relations hold:
\[ S_0=S_2,\ S_2=\tfrac{1}{2} \big( S_1+S_3 \big),\, S_1\not= S_3,\
S_3=\tfrac{1}{2} \big( S_4+S_5 \big),\, S_4\not= S_5.\] 
This implies, in particular, that the market is arbitrage-free, and,
due to its incompleteness, the set of equivalent martingale measures
is larger that just a singleton. Next, we consider a family $\{f_a\}_{a>0}$ of 
contingent claims
defined by
\[ 
f_a(\omega)=\begin{cases}
0, & \omega=\omega_1,\omega_2,\\
a, & \omega=\omega_3, \omega_4.\\
\end{cases}
\]
We are going to compare $\rho(f_a; 1)$ and
$\rho(f_a;2)$ where $\rho(f_a;t)$, $t=1,2$, is the value of the entropic
risk measure  (as
defined in (\ref{equ:def-entr}) above) of the contingent claim $f_a$, seen as
time-$t$ random variable (note that $f_a$ is $\FF_1$-measurable, for
all $a$).

Let us first focus on $\rho(f;2)$.
The set of all
martingale measures  is given by $\MM=\sets{\QQ^{\nu}}{
  \nu \in (-\tfrac{1}{6},\tfrac{1}{3})}$, where
\[ \QQ^{\nu}(\omega)=\begin{cases}
\tfrac{1}{3}- \nu, & \omega=\omega_1,\\
\tfrac{1}{3}+2 \nu, & \omega=\omega_2,\\
\tfrac{1}{2} \big( \tfrac{1}{3}- \nu \big), & \omega=\omega_3, \omega_4.\\
\end{cases}
\]
By a finite-dimensional analogue of (\ref{equ:El-Karoui}), 
 we have
\begin{equation}%
\label{equ:nu-two}
    \begin{split}
\rho(f_a;2)=\sup_{\nu\in (-1/6, 1/3)} \Big( \EE^{\QQ^{\nu}}[-f_a]-h_2(\nu)
\Big)=\sup_{\nu\in (-1/6, 1/3)} \Big( -a (1/3-\nu) - h_2(\nu) \Big),\,
    \end{split}
\end{equation}
where, as one can easily check, the relative-entropy function $h_2$ is
given by \[h_2(\nu)=\bar{h}_2(\nu)-\inf_{\mu} \bar{h}_2(\mu),\] where
\begin{multline*}
  \bar{h}_2(\nu) = \tfrac{\QQ^{\nu}[\omega_1]}{\PP[\omega_1]}
  \ln\left(\tfrac{\QQ^{\nu}[\omega_1]}{\PP[\omega_1]}\right)+
  \tfrac{\QQ^{\nu}[\omega_2]}{\PP[\omega_2]}
  \ln\left(\tfrac{\QQ^{\nu}[\omega_2]}{\PP[\omega_2]}\right)
  +\tfrac{\QQ^{\nu}[\omega_3]}{\PP[\omega_3]}
  \ln\left(\tfrac{\QQ^{\nu}[\omega_3]}{\PP[\omega_3]}\right)+
  \tfrac{\QQ^{\nu}[\omega_4]}{\PP[\omega_4]}
  \ln\left(\tfrac{\QQ^{\nu}[\omega_4]}{\PP[\omega_4]}\right).
\end{multline*}
    Similarly, 
\begin{equation}
\label{equ:nu-one}
    \begin{split}
 \rho(f_a;1)=\sup_{\nu\in (-1/6, 1/3)} \Big( \EE^{\QQ^{\nu}}[-f_a]-h_1(\nu)
\Big)=\sup_{\nu\in (-1/6, 1/3)} \Big( -a (1/3-\nu) - h_1(\nu) \Big),
    \end{split}
\end{equation}
where  the function $h_1$ is given by
 $h_1(\nu)=\bar{h}_1(\nu)-\inf_{\nu} \bar{h}_1(\nu)$, with 
\begin{multline*}
  \bar{h}_1\left(\nu\right) =
  \QQ^{\nu}[\omega_1]\ln\left(\tfrac{\QQ^{\nu}[\omega_1]}{\PP[\omega_1]}\right)+
  \QQ^{\nu}[\omega_2]\ln\left(\tfrac{\QQ^{\nu}[\omega_2]}{\PP[\omega_2]}\right)
  +\left( \QQ^{\nu}[\omega_3]+\QQ^{\nu}[\omega_4]\right) \ln
  \left(\tfrac{\QQ^{\nu}[\omega_3]+\QQ^{\nu}[\omega_4]}{\PP[\omega_3]+\PP[\omega_4]}\right).
    \end{multline*}
The expressions
(\ref{equ:nu-two}) and (\ref{equ:nu-one}) can be seen as
the Legendre-Fenchel transforms of the translated entropy functions
$h_2(1/3-\nu)$ and $h_1(1/3-\nu)$.   Therefore, by the 
bijectivity of these  transforms and the convexity of
the functions $h_1$ and $h_2$, 
the equality
$\rho(f_a;1)=\rho(f_a;2)$, for all $a>0$, would imply that
$h_1=h_2$. It is
now a matter of a straightforward  computation to show that that is, in
fact, not the case. Thus, the two  values do not  coincide, i.e., 
for at least one $a>0$,
\[ \rho(f_a;1)\not= \rho(f_a;2).\]

\paragraph{\bf b) A non-compliance example in a diffusion market
  model}\ \\
We consider a   financial market  as in section \ref{sec:general},
with $k=1$ (one risky asset) and $d=2$ (two driving Brownian motions).
It will be enough to consider a stock price process with stochastic
volatility of the form \begin{equation}
   \label{equ:stochastic-volatility}
   \begin{split}
 dS_s &= S_s( \mu\, ds+\sigma(\tB_s)\, dW^1_s), \\
 d\tB_s& = dB_s,
   \end{split}
\end{equation}
$s\geq 0$,  on an augmented filtration generated by two
independent Brownian motions $W^1$ and $W^2$, where $B=\rho
W^1+\sqrt{1-\rho^2} W^2$ is a Brownian motion correlated with
$W^1$, with the correlation coefficient $\rho\in (0,1)$.  It will
be convenient to introduce the market price of risk
$\ld(y)=\mu/\sigma(y)$, assuming throughout that $\ld:\R\to(0,\infty)$
is a strictly increasing $C^{1}$-function with
range of the form $(\eps,M)$ for some constants $0<\eps<M<\infty$.
 The trading starts at
time $t$, after which two maturities $T,\bar{T}$, 
with $T<\bar{T}$, are chosen. 

Let $C_T=-B_T$ model the payoff of a contingent claim which is,
clearly, nonreplicable.  The value of the time-$t$ entropic
($\gamma=1$) risk measure $\rho_t(C_T;T)$ equals the  
indifference
price $\nu_t(-C_T;T)$ of the claim $B_T$ measured on the trading
horizon $[t,T]$.  According to \cite{SirZar05}, $\rho_t(C_T;T)$ admits
a representation in terms of a solution to a partial differential
equation. More precisely,  taking into
 account the fact that neither the payoff $C_T$ nor the dynamics of
the volatility depend on the stock price, we have
$\rho_t(C_T;T)=p(t,-\tB_t)$, a.s.,  
where the function $p:[0,T]\times \R \to \R$ is
a classical solution of the quasilinear equation
\begin{equation}%
\label{equ:p}
    \begin{split}
\left\{ \begin{array}{l}
p_t+\sL^f p+ \tot (1-\rho^2) p_y^2=0 \\
p(T,y)=y,
\end{array}
\right.
\end{split}
\end{equation}
where $\sL^f p=
\tot p_{yy}+ \big(f_y/f-\rho \ld(y)\big) p_y$.
The function $f:[0,T]\times \R\to\R$ is the unique 
solution to the linear problem
\begin{equation}%
\label{equ:f}
    \begin{split}
\left\{  \begin{array}{l}
    f_t+\AA f =0  \\
f(T,y)=1,
  \end{array}\right.
        \end{split}
\end{equation}
 where 
$\AA f = \tot f_{yy} -\rho \ld(y) f_y -
\tot (1-\rho^2) \ld^2(y) f$.
Standard arguments show that $f$ is of class $C^{1,3}$  
and admits a representation 
in the manner of Feynman and Kac as
\begin{equation}%
\label{equ:Feynman-Kac}
    \begin{split}
f(t,y)= \EE[ e^{\int_t^T \frac{(1-\rho^2)}{2} \ld^2(Y_s)\, ds}|Y_t=y],\
(t,y)\in [0,T]\times \R,
        \end{split}
\end{equation}
where $\{Y_s\}_{s\in [t,\infty)}$ 
is the unique strong solution to 
$dY_s= dB_s-\rho \ld(Y_s)\, ds$, $Y_t=y$. In particular, there exists a
constant $C>1$ such that $1\leq f(t,y)\leq C$, for  $(t,y)\in
[0,T]\times \R$. 

Similarly, the indifference price $\nu_t(-C_T;\bar{T})$ (which equals
the value $\rho_t(C_T;\bar{T})$ 
of the maturity-$\bar{T}$ 
entropic risk measure $\rho_t(\cdot;\bar{T})$ applied to the  same
contingent claim, only on the longer horizon $[0,\bar{T}]$,  
$\bar{T}>T$) 
can be represented via $\bar{p}(t,y)$, where
$\bar{p}$ solves
\begin{equation}%
\label{equ:p-bar}
    \begin{split}
\left\{\ \begin{array}{l}
\bar{p}_t+\sL^{\bar{f}} \bar{p}+ \tot (1-\rho^2) \bar{p}_y^2=0 \\
\bar{p}(T,y)=y.
\end{array}
\right.
\end{split}
\end{equation}
Herein,  $\sL^{\bar{f}}$ is given as in (\ref{equ:f}) with $f$ replaced
by  the function 
 $\bar{f}$ which solves 
\begin{equation}%
\label{equ:f-bar}
    \begin{split}
\left\{ 
  \begin{array}{l}
    \bar{f}_t+\AA \bar{f}=0,  \\
    \bar{f}(\bar{T},y)=1.
  \end{array}
\right.
        \end{split}
\end{equation}
Just like $f$, the function $\bar{f}$ admits  
a representation analogous to (\ref{equ:Feynman-Kac})
and a uniform bound $1\leq \bar{f}(t,y)\leq \bar{C}$, for $(t,y)\in
[0,T]\times \R$. 

The goal of this example is to show that the indifference prices
$\nu(B_T;T)$ and $\nu(B_T; \bar{T})$, or, equivalently, 
 the entropic risk measures
$\rho_t(C_T;T)$ and $\rho_t(C_T;\bar{T})$, do {\em not} always coincide, i.e.,
that $p(t,y)$ and $\bar{p}(t,y)$ differ for at least one
choice of $(t,y)\in [0,T)\times\R$.
We start with an
auxiliary result, namely, 
\begin{equation}%
\label{equ:claim-one}
    \begin{split}
\frac{f_y(T,y)}{f(T,y)}\not= \frac{\bar{f}_y(T,y)}{\bar{f}(T,y)},\ \text{
  for each } y\in\R.
    \end{split}
\end{equation}
In order to establish \eqref{equ:claim-one}, we note that the function
$g:[0,T]\times \R\to \R$, defined by $g=f_y$, is a classical solution
to 
\begin{equation}%
\label{equ:g}
    \begin{split}
\left\{ 
  \begin{array}{l}
    g_t+\BB g=0  \\
    g(T,y)=0,
  \end{array}
\right.
        \end{split}
\end{equation}
where 
\[\BB g= \tot g_{yy}-\rho\ld(y) g_y - A(y) g - B(t,y),\]
with $ 
A(y)= \rho\ld'(y)+\tot(1-\rho^2)\ld^2(y)\text{ and }B(t,y)=(1-\rho^2)
\ld(y) \ld'(y) f(t,y)$. 

Thanks to the assumptions placed on $\rho$ and
$\ld$, and the positivity of  $f$,  
we have 
\begin{equation}%
\label{equ:A-B-positive}
    \begin{split}
A(y)>0\text{ and } B(t,y)>0,\text{ for all } (t,y)\in [0,T]\times\R. 
    \end{split}
\end{equation}

The function
$\bar{g}=\bar{f}_y$ is defined in an analogous fashion (only on the larger domain
$[0,\bar{T}]\times \R$) and a similar set of
properties can be derived. Since $f_y(T,y)=0$ for all $y\in\R$, it
will be enough to show that $\bar{f}_y(T,y)>0$ for all $y\in\R$. This
follows immediately from the  strict inequalities in 
 \eqref{equ:A-B-positive} and the 
Feynman-Kac representation
\begin{equation}%
\label{equ:F-K-derivative}
    \begin{split}
\bar{g}(T,y)=\bar{f}_y(T,y)= \EE[\int_T^{\bar{T}} B(t,Y_t) 
e^{\int_t^{\bar{T}} A(Y_s)\, ds}\, dt|Y_T=y],\
y\in \R.
    \end{split}
\end{equation}

Having established \eqref{equ:claim-one}, we conclude that, thanks to
the smoothness of the functions $f$ and $\bar{f}$, the operators $\sL^f$
and $\sL^{\bar{f}}$ differ in the $\tpds{y}$-coefficient in some
open neighbourhood $\NN$ of the line $\set{T}\times \R$ in $[0,T]\times \R$. 
Assuming
that $\bar{p}$ and $p$ coincide in $\NN$, subtracting the equations 
\eqref{equ:p} and
\eqref{equ:p-bar} yields
\begin{equation}
    \begin{split}
\left(\frac{f_y}{f}(t,y)-\frac{\bar{f}_y}{\bar{f}}(t,y)\right) 
\bar{p}_y(t,y)=0,\text{
  for } (t,y)\in \NN. 
    \end{split}
\end{equation}
Equation \eqref{equ:claim-one} now implies that $\bar{p}_y=0$ on
$\NN$, which is clearly in contradiction with the terminal condition
$\bar{p}(T,y)=y$, $y\in\R$. Therefore, there exists $(t,y)\in
\NN\setminus \set{T}\times \R\subseteq [0,T)\times \R$ such that
$p(t,y)\not = \bar{p}(t,y)$.

\section{Forward Entropic Risk Measures (FERM)}

In the previous section, we saw three examples of risk measures 
and their dependence 
on the specific choice of the maturity date. 
In particular, we pointed out
that the super-hedging risk measure  in \ref{sss:super-hedging}, 
as well as the  ones
 constructed in \ref{sss:closed-example}, for the class of
closed markets,  are maturity-independent. However, both  these
classes 
are rather  restrictive. Indeed, the one associated with super-hedging
is prohibitively conservative,
 while the other  requires the
rather stringent assumption of market closedness.

In this section, we introduce a new family of convex risk measures
that  have
the \textbf{maturity independence} property and, at the same time,  are
applicable to a wide range of settings. Their construction is based on
the idea mentioned in the introductory paragraph of Subsection
 \ref{sss:no-hor-ind}, but avoids the pitfalls responsible for the
 failure of examples a) and b) following it. 

 The risk measures we are going to introduce are closely related to
 indifference prices. The novelty of the approach is that the
 underlying risk preference functionals are not tied down to a
 specific maturity, as it has been the case in the standard expected
 utility formulation. Rather, they can be seen as specified at
 initiation and subsequently ``generated'' across all times.  This
 approach was proposed by the first author and M. Musiela (see
 \cite{MusZar03, MusZar05, MusZar06a, MusZar06,MusZar07}) and is briefly
 reviewed below.

\subsection{Forward exponential performances}
The notion of a forward performance process has arisen from the 
search for  ways to measure the
performance of investment strategies across {\em all} times in $[
0,\infty)$. 
In order to produce a nontrivial such object, we look for a random
field $U=U_t(\omega,x)$
defined for 
\textit{all }times $t\geq 0$ and
 parametrized by a wealth argument $x$ such that the mapping $x\mapsto
 U_t(\omega,x)$ admits the classical properties of utility
 functions. More precisely, we have the following definition:
\begin{definition}
\label{def:utility-random-field}
A mapping $U:[0,\infty)\times\Omega\times\R\to\R$ is called a
\define{performance random field} if
\begin{enumerate}
\item for each $(t,\omega)\in [0,\infty)\times \Omega$, 
the mapping $x\mapsto
U_t(x,\omega)$ defines a utility function: it is strictly concave,
strictly increasing, continuously differentiable and 
 satisfies the Inada conditions $\lim_{x\to\infty}
U'(x)=0$ and $\lim_{x\to -\infty} U'(x)=+\infty$,
\item  $U_{\cdot}(\cdot,\cdot)$ is measurable with respect
to the product of the 
progressive $\sigma$-algebra on $\Omega\times [0,\infty)$ and the
Borel $\sigma$-algebra on $\R$, and 
\item $\EE \abs{U_t(x)} <\infty$, for all $(t,x)\in
[0,\infty)\times \R$. 
\end{enumerate}
\end{definition}
\begin{remark}\ 
\begin{enumerate}
\item The last requirement in Definition
  \ref{def:utility-random-field} implies, in particular, that
$\EE \abs{U_t(\xi)} <\infty$, for all random variables
$\xi\in\linf$. 
\item
It is possible to construct a parallel theory where the performance
functions $U_t(\omega,\cdot)$ are defined on the positive semi-axis
$(0,\infty)$. We choose the domain $\R$ for the wealth argument $x$
because it leads to a slightly simpler analysis, and because the
examples to follow  will be based on the exponential function.
\end{enumerate}
\end{remark}

On an
arbitrary trading horizon, say $[s,t]$, $0\leq s<t<\infty$, 
 the investor whose preferences are described by the  random
 field $U$  seeks to 
maximize the expected investment performance: 
\begin{equation}
\label{equ:def-v}
V_{s}(x) =\esssup_{\pi\in \AAb }\EE[
U_{t}( X_{t}^{x,\pi }) | \FF_s ],\ 0\leq s\leq t.  
\end{equation}%
Herein, $X^{x,\pi}$ 
denotes the investor's wealth process, $x\in\R$ the investor's initial
wealth at time $s$,
 and  $\pi $ a generic investment strategy belonging to $\AAb$
(the set of admissible policies introduced in Subsection
\ref{sss:portfolio-proc}.) 
To concentrate on the new notions, we abstract throughout from control and
state constraints, as well as the most general specification
of admissibility requirements. 

It has been argued in \cite{MusZar07} that the class of performance random
fields with the additional property 
\begin{equation}
\label{equ:V-U}
V_{t}( x) = U_{t}( x),\text{ a.s.} \ \ \forall\, t\in [0,\infty),\ x\in\R,
\end{equation}%
possesses several desirable properties and gives rise to an
analytically tractable theory.
\begin{definition}
A random field $U$ satisfying
(\ref{equ:V-U}), where $V$ is defined by (\ref{equ:def-v}),
 is called \define{self-generating}.
\end{definition}
\begin{remark}
\label{rem:classical}
We remind the reader that a classical example of a self-generating
performance random field (albeit {\em only on the finite horizon $[0,T]$}) 
is the traditional value function, defined as
\[
U_{t}(x) =\esssup_{\pi\in\AAb}
\EE[ 
U_{T}( X_{T}^{x,\pi })| \mathcal{F}_{t}],\ t\in [0,T],\,x\in\R,
\]%
where $T$ is a prespecified maturity beyond which no investment activity is
measured, and $U_T(\cdot,\cdot):\Omega\times\R\to\R$ is a classical
(state-dependent) utility function (see, for example, 
\citeand{KarLehShrXu90,KraSch99,KarZit03}{Zit05}).
 When the horizon is infinite, such a construction will not
produce any results. Indeed, there is no appropriate time for the final datum
to be given. 
\end{remark}

\medskip

What (\ref{equ:def-v}) and (\ref{equ:V-U}) tell us is that (under
additional regularity conditions) the sought-after
criterion (performance random field) $U$ 
must have the property that the stochastic process 
$U_{t}(\Xxp_t) $ is a
supermartingale for an arbitrary control $\pi\in\AAb$ and 
becomes ``closer and closer'' to a martingale as the controls get ``better
and better''. In the case when the class of 
control problems \eqref{equ:def-v}
actually admits an optimizer $\pi^*\in\AAb$ (or in some larger,
appropriately chosen,  class), the composition
$U_t(X^{x,\pi^*}_t)$ becomes a martingale. 

In the traditional framework,  
as already mentioned in Remark \ref{rem:classical}, 
the datum (terminal utility) is assigned at some
fixed future time $T$.  
Alternatively, in the case of an infinite time horizon, it is more
natural to think of the datum $u_0:\R\to\R$ 
as being assigned at time $t=0$, and a
self-generating 
performance random field $U_t$  chosen so that $U_0(x)=u_0(x)$. It is
because of this interpretation that the self-generating performance random
fields may, also, be referred to as \define{forward performances}. 

The notion of forward performance processes was first developed for
binomial models in \cite{MusZar03} and \cite{MusZar05} and later
generalized to diffusion models with a stochastic factor
(\cite{MusZar06}) and, more recently, to
models of It\^ o asset price dynamics (see, among
others, \cite{MusZar06} and \cite{MusZar07}, as well as
\cite{BarRogTeh07}). 
A related stochastic
optimization problem that allows for semimartingale price processes
and random horizons can be found in \cite{ChoStrLi07}. A similar
notion of utilities without horizon preference was developed in
\cite{HenHob07}; therein, asset prices are taken to be lognormal,
leading to deterministic forward solutions.
 
While traditional  performance random fields on finite horizons
are
straightforward to construct and characterize, 
producing a ``forward'' performance random field on 
$[0,\infty)$ from a given initial datum $u_0$ is considerably
more difficult. Several examples of such a construction, 
all based on the exponential
initial datum, are given in the following subsection.
These random
fields are the most important building blocks for the
 class of maturity-independent risk measures presented in
 subsection \ref{sse:FERM}, below. 
\begin{definition}
A performance random field $U$ is called a \define{forward
  exponential performance} if 
\begin{itemize}
\item[a)] it is self-generating, and 
\item[b)]  there exists a constant $\gamma>0$, such
that
 \begin{equation}
U_{0}(x) =-e^{- \gamma x},\ x\in \R.
\label{equ:forw-exp}
\end{equation}
\end{itemize}
\end{definition}
The construction presented below 
can be found in \cite{MusZar06a}.  The
 assumptions and definitions from section \ref{sec:general} will be
 used in the sequel without explicit mention. 
\begin{theorem}[Theorem 4 in \cite{MusZar06a}]
\label{thm:yza}
Let $\prfi{Y_t},\prfi{Z_t}$ be two continuous 
processes solving 
\begin{equation}
dY_{t}=Y_{t}\delta _{t} ( \lambda_{t}dt+dW_{t})  \label{equ:Y}
\end{equation}%
and 
\begin{equation}
dZ_{t}=Z_{t}\phi _{t} dW_{t},  \label{equ:Z}
\end{equation}%
with $Y_{0}=1/\gamma>0$, $Z_{0}=1$, for a fixed, but arbitrary
$k$-dimensional coefficient processes $\prfi{\delta_t} $ and
$\prfi{\phi_t}$ which are assumed to be $\FFF$-adapted, and  
 with $\delta$ satisfying
$\sigma_t \sigma^{+}_t\delta_t =\delta_t$, for all $t\geq 0$, a.s.
We, also, assume that $\delta$ and $\phi$ are regular enough for the
integrals in (\ref{equ:Y}) and (\ref{equ:Z}) to be well defined, and
that,
 when restricted to any finite interval $[0,t]$,
the process $Z$ is a positive martingale, and $Y$ is uniformly bounded from
above and away from zero. 

Let the process $\prfi{A_t}$ be defined as %
\begin{equation}
A_{t}= \int_{0}^{t}\big\| \sigma _{s}\sigma _{s}^{+}( \lambda
_{s}+\phi _{s}) -\delta _{s}\big\|^{2}ds.  \label{equ:A}
\end{equation}%
Then, the random field $U$, given by 
\begin{equation}
U_{t}( x;\omega ) =-Z_{t}\exp\left( -\frac{x}{Y_{t}}%
+\frac{A_{t}}{2}\right),   
\label{equ:form-entr-forw}
\end{equation}%
is a forward exponential performance. In particular, 
for $0\leq s\leq t$ and $\xi \in 
\mathbb{L}^{\infty }( \mathcal{F}_{s}) $, we have
\begin{equation}
  U_{s}( \xi ) =\esssup_{\pi \in \mathcal{A}_{bd}}
  \EE\left[ \left. U_t\left(\xi +\int_{s}^{t}\pi_{u}dS_{u}\right) 
\right|\FF_s\right], \text{ a.s.}
\label{equ:forw-semi}
\end{equation}
\end{theorem}

\begin{remark}
\label{rem:disc-entr-forw}
In \eqref{equ:form-entr-forw} above, one can give a natural financial
interpretation to the processes $Y$ (which normalizes the wealth
argument) and $Z$ (which appears as a multiplicative factor). One
might think of $Y$ as a benchmark (or a num\' eraire) in
relation to which we wish to measure the performance of our investment
strategies.  The values of the 
process $Z$,
on the other hand, can be thought of as Radon-Nikodym derivatives of
the investor's 
subjective probability measure with respect to the measure $\PP$. 
\end{remark}
\subsection{Forward entropic risk measures}
\label{sse:FERM}
We are now ready to introduce the \textit{forward entropic risk
  measures} (FERM).
We start with an auxiliary object, denoted by $\rho(C;t)$.
\begin{definition}
\label{def:for-exp-per}
Let $U$ be the forward exponential performance  
defined in (\ref{equ:form-entr-forw}), and let $t\geq 0$ be arbitrary,
but fixed. For a contingent claim written at time $s=0$
and yielding a payoff $C\in \linf(\FF_t)$, we define 
$\rho(C;t)\in\R$ as the unique solution of
\begin{equation}%
\label{equ:indif-pric}
    \begin{split}
\sup_{\pi\in\AAb}
\EE\Big[ U_{t}\Big( x&+\int_{0}^{t}\pi_{s}dS_{s}\Big) \Big]
=
\sup_{\pi \in \AAb}
\EE\Big[ U_{t}\Big( x+\rho( C;t) + C+\int_{0}^{t}\pi_{s}dS_{s}\Big) \Big],\ \forall\, x\in\R.
    \end{split}
\end{equation}
The mapping $\rho(\cdot;t):\linf(\FF_t)\to\R$ is called the
\define{$t$-normalized forward entropic measure}. 
\end{definition}
One can, readily, check that the equation (\ref{equ:indif-pric}) indeed
admits a unique solution (independent of the initial wealth $x$),
 so that the $t$-normalized forward entropic
measures are well defined.  The reader can convince
him-/herself of the validity of the following result:
\begin{proposition}
\label{pro:t-norm-prop}
The $t$-normalized forward entropic risk measures are
re\-pli\-ca\-ti\-on-in\-va\-ri\-ant convex risk measures on
$\linf(\FF_t)$,  
for each $t\geq 0$. 
\end{proposition}

The {\em fundamental} property in which forward entropic risk measures
differ from a generic replication-invariant risk measure (see examples
in Subsection \ref{sss:closed-example}) is the following:

\begin{proposition}
\label{pro:price-the-same}
For $0\leq s<t<\infty$, and $\Cs\in\linf(\FF_s)$,
consider the $s$- and $t$-normalized forward entropic measures
$\rho( \Cs;s)$ and $\rho(\Cs;t)$ applied to the contingent 
claim $\Cs$. Then,%
\begin{equation}
\rho(\Cs;s)=\rho(\Cs;t).
\label{equ:hor-inv-sim}
\end{equation}%
More generally, for $\Csr\in \linf(\FF_{r})$, where $0\leq r<s<t<\infty$, we
  have
\begin{equation}
\rho(\Csr;s)=\rho(\Csr;t).
\label{equ:hor-inv-dou}
\end{equation}
\end{proposition}
\begin{proof}
We are only going to establish (\ref{equ:hor-inv-sim}) since (\ref%
{equ:hor-inv-dou}) follows from similar arguments. 
To this end, 
note that a self-financing policy $\pi \in \mathcal{A}_{bd}$ if and only
if $\pi \mathbf{1}_{[ 0,t] }\in \mathcal{A}_{bd}$ and $\pi 
\mathbf{1}_{( t,\infty ) }\in \mathcal{A}_{bd}$. 
Using Definition \ref{def:for-exp-per} at  $x=0$, we obtain
\begin{equation}
    \nonumber 
    \begin{split}
U_{0}(0) 
&=\sup_{\pi\in\AAb}\EE[ U_t(\rho(\Cs;t)+\Cs+\int_0^t \pi_u\,\, dS_u)]\\
&=\sup_{\pi,\pi' \in\AAb} \EE\left[ \EE[ U_t( \rho(\Cs;t)+\Cs+\int_0^s \pi_u\,
dS_u+\int_s^t \pi'_u\, dS_u)|\FF_s]\right] \\
&=\sup_{\pi\in\AAb} \EE\left[ \esssup_{\pi'\in\AAb} 
\EE[ U_t( \rho(\Cs;t)+\Cs+\int_0^s \pi_u\,
dS_u+\int_s^t \pi'_u\, dS_u)|\FF_s]\right] \\
&=\sup_{\pi\in\AAb} \EE\left[ U_s( \rho(\Cs;t)+\Cs+\int_0^s \pi_u\,
dS_u)\right],  \\
    \end{split}
\end{equation}
where we used the semigroup property \eqref{equ:forw-semi} of $U$ and
the fact that the random variable 
$\rho(\Cs;t) +\Cs+\int_0^s \pi_u\, dS_u$ is an element of $\linf(\FF_s)$,
for all $\pi\in\AAb$. We compare the obtained expression with the
defining equation \eqref{equ:indif-pric} to conclude that
$\rho(\Cs;t)=\rho(\Cs;s)$. 
\end{proof}

We  are now ready to define the forward entropic risk measures:
\begin{definition}
\label{def:main}
For $C\in \LL$, define the \define{earliest maturity} 
$t_{C}\in [0,\infty)$ \define{of} $C$ as
\begin{equation}
t_{C}=\inf \sets{t\geq 0}{C\in \FF_t}.  \label{equ:t(C)}
\end{equation}%
The \define{forward entropic risk measure} $\nu:\LL\to\R$ 
is defined as 
\begin{equation}
\rho(C) =\rho(C;t_{C}) ,
\label{equ:FERM-def}
\end{equation}%
where $\rho( C;t_{C}) $ is the value of the 
$t_{C}$-normalized forward
entropic risk measure, defined in (\ref{equ:indif-pric}),
  applied to the contingent claim $C$.
\end{definition}
The focal point of the present section is the following theorem:
\begin{theorem}
\label{thm:main}
The mapping $\rho:\LL\to\R $ is a maturity-independent risk measure.
\end{theorem}
\begin{proof}
We need to verify  the axioms (1)-(4) of  Definition 
\ref{def:horiz-indep}. 
Axioms (1) and (3) follow directly from elementary 
 properties of the
$t$-normalized forward risk measures.
To show axiom $(2)$ we take $\lambda \in ( 0,1) $
and $C_{1},C_{2}\in \LL$. Then, since $\ld C_1+(1-\ld)C_2 \in
\FF_{\max(t_{C_1},t_{C_2})}$, we have 
$\max (t_{C_{1}},t_{C_{2}})\geq t_{\lambda C_{1}+(
1-\lambda ) C_{2}}$. 
Therefore,  
\begin{equation}
    \nonumber 
    \begin{split}
\rho( \lambda C_{1}+( 1-\lambda ) C_{2}) 
&=\rho( \lambda C_{1}+( 1-\lambda ) C_{2};t_{\lambda C_{1}+( 1-\lambda ) C_{2}})\\ 
&=\rho( \lambda C_{1}+( 1-\lambda ) C_{2};\max(t_{C_1},t_{C_2}) ) ,
    \end{split}
\end{equation}
where we used (\ref{equ:hor-inv-sim}). Using  property
\eqref{equ:hor-inv-sim} and  
the fact that  the $t$-forward
entropic risk measures are convex risk measures, 
we get
\begin{equation}
    \nonumber 
    \begin{split}
\rho( \lambda C_{1}+( 1-\lambda ) C_{2}; ) 
& \leq
\lambda \rho( C_{1};\max(t_{C_1},t_{C_2})) +
( 1-\lambda ) \rho( C_{2};\max(t_{C_1},t_{C_2}))\\
&=  \lambda \rho( C_{1};t_{C_1}) +
( 1-\lambda ) \rho( C_{2};t_{C_2} )\\
&= \lambda \rho(C_1)+(1-\ld)\rho(C_2).
    \end{split}
\end{equation}

It remains to check  the replication and maturity independence axiom (4).
 To this end, we let $\xi =\int_{0}^{\infty }\pi_{u}dS_{u}$ for 
 some portfolio process $\pi \in \AAb$. We need to show that
\[
\rho( C+\xi ) =\rho( C), 
\]%
for any $C\in\LL$. 
Observe that $\max ( t_{C},t_{\xi }) \geq t_{C+\xi }$ and,
therefore, by (\ref{equ:hor-inv-sim}) and (\ref{equ:FERM-def}), we have 
\begin{equation}
    \nonumber 
    \begin{split}
\rho( C+\xi ) =\rho( C+\xi ;t_{C+\xi }) 
=\rho( C+\xi ;\max ( t_{C},t_{\xi }) ). 
    \end{split}
\end{equation}
On the other hand, Proposition \ref{pro:t-norm-prop},
the form of $\xi $ and (\ref{equ:hor-inv-sim})  yield 
\[
\rho( C+\xi ;\max ( t_{C},t_{\xi }) ) =\rho( C;\max ( t_{C},t_{\xi }) )
=\rho(C;t_C)=\rho(C),
\]
establishing axiom (4).
\end{proof}

Next, we provide an explicit representation of the forward entropic risk
measures.
\begin{theorem}
\label{thm:explicit-ferm}
Let $Y,Z$, $A$ and $U_t(\cdot)$ be as in Theorem \ref{thm:yza}. For 
$C\in\LL$, its forward entropic risk measure is given by
\begin{equation}
    \label{equ:explic}
    \begin{split}
\rho(C)= 
\inf_{\pi\in\AAb}\Big( \frac{1}{\gamma} 
\ln \EE[- U_t(C+\int_0^t \pi_s\
dS_s)]\Big),\ \text{ for any } t\geq t_C,
    \end{split}
\end{equation}
where $t_C$ is defined in (\ref{equ:t(C)}). 
\end{theorem}
\begin{proof}
Equation 
(\ref{equ:indif-pric}) (with 
$x=-\rho(C;t)$ and  $t\geq t_C$) and  
the property (\ref{equ:forw-semi})
of the random field $U$, yield that 
\begin{equation}%
\label{equ:aux-expr}
    \begin{split}
-\exp(\gamma \rho(C)) = \sup_{\pi\in\AAb} \EE[ U_t( C+\int_0^t
\pi_s\, dS_s)],\text{ for any $t\geq t_C$.}
    \end{split}
\end{equation}
By (\ref{equ:forw-semi}), the right-hand side of (\ref{equ:aux-expr})
is independent of $t$ for $t\geq t_C$. 
\end{proof}
\subsection{Relationship with dynamic risk measures}
\label{sse:hi-dyn}
Before we present concrete examples of maturity-independent risk
measures in section  \ref{sec:examples}, let us briefly discuss  their
relationship with the dynamic risk measures (see the introduction for
references). A family of mappings
$\rho_s(\cdot;t):\linf(\FF_t)\to\linf(\FF_s)$, where $0\leq
s\leq t\leq T$, with $T\in [0,\infty]$,
 is said to be a \define{dynamic (time-consistent) risk measure} 
if each $\rho_s(\cdot;t)$ satisfies the analogues of
the axioms of convex risk measures and the semi-group property 
\[ \rho_s(-\rho_t(f;u);t)=\rho_s(f;u),\ 0\leq s\leq t\leq u\leq T,\]
holds. Using a version of Definition \ref{def:main} and Theorem
\ref{thm:main}, the reader can readily check that each
replication-invariant dynamic risk measure defined on the whole
positive semi-axis $[0,\infty)$ (i.e., when $T=\infty$) gives rise to
a maturity-independent risk measure. Under certain conditions, the
reverse construction can be carried out as well (details will be
presented in \cite{Zit08}).

The philosophies of the two approaches are quite different,
though. Perhaps the best way to illustrate this point is through the
analogy with the expected utility theory. Dynamic risk measures
correspond to the traditional  
utility framework where a system of
decisions relating various 
maturity dates is interlaced together
through a consistency criterion. The maturity-independent risk measures
take the opposite point of view and correspond to forward
performances. While the dynamic risk measures are natural in the case
$T<\infty$, the maturity-independent risk measures fit well with
infinite or un-prespecified maturities.

\section{Examples}
\label{sec:examples}
In this section, we provide two representative classes of forward entropic
risk measures. 
First, we single out some of the special cases obtained when specific
choices for the processes $Z$ and $Y$ (of Definition
\ref{thm:yza}) are used in conjunction with  Definition \ref{def:main}  of the
forward entropic risk measures.
Then, we illustrate the versatility of the general notion of 
maturity-independent risk measures by constructing an example in an incomplete
binomial-type model.
Even though the remainder of the
paper is set 
in an
It\^{o}-process model framework, the latter example is not.
The reader can easily translate all the
relevant definitions and results to fit this model. Some background
and technical details pertaining to this example can be found in
\cite{MusSokZar07}. 
\subsection{It\^o-process-driven markets}
This example is set in a financial market described in section 
\ref{sec:general}, with $k=1$ (one risky asset) and $d=2$ (two driving
Brownian motions). Without loss of generality, we assume that
$\sigma^{12}_t\equiv 0$, and $\sigma_t=\sigma^{11}_t>0$,  i.e., 
that the second Brownian motion
does not drive the tradeable asset. In this case, we have
$\ld_t=(\ld^1_t,\ld^2_t)$, where 
$\ld^1_t=\mu_t/\sigma_t$ and $\ld_t^2=0$.
Therefore, the stock-price process satisfies
\[ dS_t=S_t( \mu_t\, dt+\sigma_t\, dW^1_t),\]
on an augmented filtration generated by a $2$-dimensional Brownian
motion $(W^1,W^2)$. 
The processes $Z,Y,A$ from
Theorem \ref{thm:yza} can be written as
\begin{equation}
dY_{t}=Y_{t}\delta _{t} ( \lambda^1_{t}dt+dW^1_{t}),\ Y_0=1/\gamma>0, \ 
dZ_{t}=Z_{t}\phi _{t} dW^1_{t},\ Z_0=1,  \label{equ:YeZe}
\end{equation}%
and 
\begin{equation}
  A_{t}= \int_{0}^{t}( \lambda^1_s+\phi _{s} -\delta _{s})^{2}ds, \
  A_0=0,  
\label{equ:A-new}
\end{equation}%
subject to a choice of two processes $\phi$ and $\delta$, under the
regularity conditions stated in  Theorem \ref{thm:yza}.

\medskip

\paragraph{\bf a) $\phi\equiv\delta\equiv0$.}\ \ 
In this case, $Z_t\equiv 1$, $Y_t\equiv 1/\gamma$, 
$A_t\equiv \int_0^t (\ld^1_s)^2\, ds$ and the random  field
$U$ of \eqref{equ:form-entr-forw} becomes
\[ U_t(x)=-\exp(-\gamma x+\frac{A_t}{2} ).\]
Using the indifference-pricing 
equation \eqref{equ:indif-pric} and the self-generation property 
\eqref{equ:forw-semi} of $U_t$, 
we deduce that for $C\in\LL$, the value $\rho(C)$ satisfies
\begin{equation}
    \nonumber 
    \begin{split}
-\exp(\gamma \rho(C))= 
\sup_{\pi\in\AAb}
\EE\left[-\exp\left(- \gamma (C+\int_0^t \pi_s\, dS_s) 
+\frac{A_t}{2}\right)\right],\ 
\text{ for any } t\geq t_C.
    \end{split}
\end{equation}
On the other hand, the classical (exponential) 
indifference price, 
$\nu(C-\tfrac{A_t}{2\gamma} ;t)$, of the contingent claim
$C-\tfrac{A_t}{2\gamma}$, maturing at time $t$, satisfies
\begin{equation}
   \nonumber 
   \begin{split}
&\sup_{\pi\in\AAb} \EE[-\exp(-\gamma(\nu(C-\tfrac{A_t}{2\gamma};t)
+\int_0^t \pi_s\, dS_s ))]=\\
=&\sup_{\pi\in\AAb} \EE[
-\exp(-\gamma(C-\tfrac{A_t}{2\gamma}+\int_0^t \pi_s\, dS_s ))].
   \end{split}
\end{equation}
With $H_t=\ln \sup_{\pi\in\AAb} \EE[-\exp(-\gamma \int_0^t \pi_s\, dS_u)]$
(which will be recognized by the reader familiar with exponential
utility maximization as the aggregate relative entropy), we now have
\begin{equation}%
\label{equ:t-wants-a-number}
    \begin{split}
\rho(C)=-\nu(C-\tfrac{A_t}{2\gamma};t)-\tfrac{1}{\gamma}H_t,\text{
  for any }t\geq t_C.
    \end{split}
\end{equation}

\medskip

\paragraph{\bf b) $\delta\equiv0$.}\ \ 
Then $Y_t\equiv 1/\gamma$, 
$A_t\equiv \int_0^t (\ld^1_s+\phi_s)^2\, ds$, and the random  field
$U$ of \eqref{equ:form-entr-forw} takes the form
\[ U_t(x)=-Z_t \exp(-\gamma x+\frac{A_t}{2}).\]
The risk measure $\rho(C)$ can be represented as
in (\ref{equ:t-wants-a-number})
above, with one important
difference. Specifically, 
the (physical) probability measure $\PP$ has to be replaced by
the probability $\tilde{\PP}$  whose Radon-Nikodym
derivative w.r.t.~$\PP$ is given by $Z_t$ on $\FF_t$, for any $t\geq
0$. 

\medskip

We leave the discussion of further examples in this setting 
- in particular for the case
$\delta\not = 0$ -  
for the upcoming work of one of the authors 
\cite{Zit08}.

\subsection{The binomial case}
Let $(\Omega,\FF,\PP)$ be a probability space on which two 
sequences 
$\sqt{\xi_t}$ and $\sqt{\eta_t}$  of random
variables are defined. 
The stochastic processes  $\sqtz{S_t}$ and $\sqtz{Y_t}$ 
are defined, in turn,  as follows:
\[S_t=\prod_{k=1}^t \xi_k,\ Y_t=\prod_{k=1}^t
\eta_k,\ \ t\in\N,\  S_0=Y_0=1.\ \]  
The process $S$ models the evolution of a (traded) risky asset, and
$Y$ is a (non-traded) factor. We assume, for simplicity, that the
agents are allowed to invest in a zero-interest riskless bond $S^0\equiv 1$. 
The following two filtrations are naturally defined 
on $(\Omega,\FF,\PP)$:
\begin{equation}
    \nonumber 
    \begin{split}
\FF^S_t &=\sigma(S_0,S_1,\dots, S_t)=\sigma(\xi_1, \dots, \xi_t),\
t\in\N_0,\text{ and }\\
\FF_t&=\sigma(S_0,Y_0,S_1,Y_1,\dots, S_t,Y_t)=
\sigma(\xi_1, \dots, \xi_t, \eta_1, \dots, \eta_t),\
t\in\N_0
    \end{split}
\end{equation}
We assume that for each $t\in \mathbb{N},$ there exist $\xi _{t}^{u},\xi
_{t}^{d},\eta _{t}^{u},\eta _{t}^{d}\in \mathbb{R}$ with $0<\xi
_{t}^{d}<1<\xi _{t}^{u}$ and $0<\eta _{t}^{d}<\eta _{t}^{u}$ such that $%
\mathbb{P}[  \xi _{t}=\xi _{t}^{u}| \mathcal{F}_{t-1}] 
=1-\mathbb{P}[ \xi _{t}=\xi _{t}^{d}| \mathcal{%
F}_{t-1}] >0$, a.s.,  and $\mathbb{P}[  \eta _{t}=\eta
_{t}^{u}] =1-\mathbb{P}[  \eta
_{t}=\eta _{t}^{d}] .$

The agent starts with initial wealth   $x\in\R$, and 
 trades in the market by holding $\alpha_{t+1}$ shares of
the asset $S$ in the interval $(t,t+1]$, $t\in\N_0$, financing his/her
purchases by borrowing (or lending to) the risk-free bond
$S^0$. Therefore, the  wealth process $\sqtz{X_t}$ is given by
\[
X_{t}=x+\sum_{k=0}^{t-1} \alpha_{k+1} (S_{k+1}-S_k),\ t\in\N,
\]%
with $X_0=x$. 
It can be shown that, for each $t\in \mathbb{N},$ there exists a unique
minimal martingale measure $\mathbb{Q}^{\left( t\right) }$ on $\mathcal{F}%
_{t}$ (see \cite{MusSokZar07} for details).

Define the $\FF_t$-predictable ($\FF_{t-1}$-adapted)
 process $\sqt{h_t}$  given by
\[
h_{t}=q_{t}\ln \left(\frac{q_{t}}{\mathbb{P}\left[ A_{t}\left\vert \mathcal{F}%
_{t-1}\right. \right] }\right)+\left( 1-q_{t}\right) \ln 
\left(\frac{1-q_{t}}{1-\mathbb{P%
}\left[ A_{t}\left\vert \mathcal{F}_{t-1}\right. \right] }\right),\text{ \ \ }%
t\in\N_0,
\]%
with
\[
A_{t}=\left\{ \omega\,:\,\xi _{t}\left( \omega \right) =\xi_{t}^{u}\right\} 
\text{ and } q_{t}=\frac{1-\xi_{t}^{d}}{\xi_{t}^{u}-\xi
_{t}^{d}}=\QQ^{(t)}\left[ A_t |\FF_{t-1}\right].\]

In \cite{MusSokZar07} (see, also, \cite{MusZar03}) it is shown that the
 random field $U:\Omega\times\N_0\times\R\to\R$ defined by
\[
U_{t}(x)=-\exp \left( -x+\sum_{k=1}^{t}h_{k}\right),
\]%
is a forward exponential performance.
We, also, consider  the inverse $U^{-1}$ of $U$ given by  
\[
U_{t}^{-1}\left( y\right) =-\ln \left( -y\right) -
\sum\limits_{k=1}^{t}h_{k}, 
\]%
for $y\in (-\infty,0)$ and  $\{h_{t}\}_{t\in\N_0}$ as above.

 For $t\in\N_0$, we define the  (single-period) 
 iterative forward price functional
 $\EN^{(t,t+1)}:\linf(\FF_{t+1})\to\linf({\FF_t})$,  given by 
\[
\EN^{(t,t+1)}(C)=\EE_{\QQ^{(t+1)}}\Big[
-U_{t+1}^{-1}\Big( \EE_{\QQ^{(t+1)}} [ U_{t+1}( -C)
| \mathcal{F}_{t}\vee \mathcal{F}_{t+1}^{S}]\Big)
\Big| \mathcal{F}_{t} \Big] , 
\]%
for any $C\in \linf(\FF_{t+1})$. Similarly, for $t<t'$ and  
$C\in\linf(\FF_{t'})$
we define the
(multi-step) forward pricing functional
$\EN^{(t,t')}:\linf(\FF_{t'})\to\linf(\FF_t)$ by 
\[
\EN^{(t,t')}(C)=
\mathcal{E}^{(t,t+1)}\Big(\mathcal{E}^{(t+1,t+2)}\big(\dots
(\mathcal{E}^{(t^{\prime }-1,t^{\prime })}(C))\big)\Big).
\]

\begin{proposition}
\label{pro:binom-example}
Let $\rho \left( \cdot \text{ };t\right) :\mathbb{L}^{\infty }\left( 
\mathcal{F}_{t}\right) \rightarrow \mathbb{R}$ be defined by 
\[
\rho \left( C;t\right) =\mathcal{E}^{(0,t)}(C). 
\]%
Then, the mapping $\rho :\mathcal{L}=\cup_{t\in\N_0}
\linf(\FF_t)\rightarrow \mathbb{R}$, defined by 
\[
\rho (C)=\rho (C;t_C ) 
\]%
for $t_C =\inf \left\{ t\geq 0:C\in \mathcal{F}_{t}\right\} $
is a maturity-independent convex risk measure.
\end{proposition}
The statement of the Proposition 
follows from an argument analogous to the one in the
proof of Proposition \ref{pro:price-the-same}. For a detailed
exposition of all steps, see \cite{MusSokZar07}.
\section{Summary and  future research} 
The goals of the
present paper are two-fold: 
\begin{enumerate}
\item to bring forth and illustrate the concept of maturity-independent
  risk measures, and
\item to provide a class of such measures. 
\end{enumerate}
 Two examples - one  defined on a finite probability
 space and the other in an It\^ o-process setting - are
  given. Their analysis  shows that, while plausible and simple from
  decision-theoretic point of view, the notion of maturity independence
  is non trivial and reveals an interesting structure .

One of the major sources of appeal of the theory of
maturity-independent risk measures is, in our opinion, the fact that it
opens a venue for a wide variety of research opportunities both from
the mathematical, as well as the financial points of view. One of these
directions, which we intend to pursue in forthcoming work 
(see \cite{Zit08}),
follows the link between maturity independence and forward performance
processes in
the direction opposite to the one explored here: while forward entropic
risk measures provide a wide class of examples of maturity-independent
risk measures, it is natural to ask whether there are any others. In
other words, we would like to give a full characterization of
maturity-independent risk measures arising from performance random fields.
Such a characterization would not only complete the outlined theory
from the mathematical point of view; it would also provide a firm
decision-theoretic foundation for the sister theory of forward
performance processes.
\def\cprime{$'$} \def\cprime{$'$}

\end{document}